\documentclass[aps, prl, reprint, twocolumn, amsmath, showkeys, showpacs]{revtex4-1}

\usepackage{graphicx}
\begin{document}

\title{Wave-like avalanche prpagation in the continuum field model of self-organized criticality}

\author{Dong Keun Oh, and Hogun Jhang}
\email[]{spinhalf@nfri.re.kr, and hgjhang@nfri.re.kr}

\affiliation{WCI Center of Plsama Theory, National Fusion Research Institute, Republic of Korea}

\date{\today}

\begin{abstract}

Travelling wave is identified as the mechanism of avalanche propagation
in the continuum SOC (self-organized critical) system. Recovering 
the hidden causality based on a generalization of Fick's law, we lead
the equivalent continuum equation which has spatiotemporal nonlocality.
Taking into account of the hyperbolicity from the retarded response
of the nonlocal kernel, it is possible to capture the propagating avalanche
in constant speed.
Verifying the computation, we analyze the evolution of instability as a
nonlinear wave under the control of the threshold dynamics,
which imposes the {\it metastability} as the survival condition of avalanche.
Being consistent with the basic assumptions of SOC theory,
the result shows unified features of the existing models.
This provides a concept toward the SOC framework
based on the physical principles of transport process.

\end{abstract}

\pacs{05.45.-a, 05.65.-a, 52.25.Fi, 64.60.Ht}
\keywords{self-organized criticality, avalanche propagation, heat waves}
\maketitle

\section{}

As a prominent theoretical framework on the scale invariance without fine tuning of
control parameters, SOC (self-organized criticality) \cite{1,2} means a conceptual state
bearing common aspects of complexity in many dissipative nonlinear systems \cite{3}.
Clearly specified by the threshold of local instability, 
SOC can be characterized by very long spatiotemporal correlation mediated by the cascade of instability
$i.e.$ avalanche propagation \cite{3,4}. 
In spite of many critical insights into the physical system dyanmically hanging
around a ``critical point'' \cite{5}, most of the generic implementations
of SOC are ``physics-free'', for instance,
represented by the evolution of digital blocks $i.e.$ cellular automation (CA) \cite{1,2,3,4}.
To associate such an abstract state with physical observables, there is a need of
deductive approach to describe ``how the avalanche works'' from the generally accepted
assumptions of SOC state. In other words, in the physical system, the SOC or
SOC-like behaviors will be properly correlated to the underlying nonlinear processes,
only if the SOC framework gives a quantitative description of avalanche
in terms of the local instability. 

Many continuum theories are usually simplification of SOC dynamics
replacing the threshold condition with a perturbative nonlinear term \cite{6}.
It's because that the complex structure of macroscopic avalanche can be hardly deduced
from the discrete relaxation rule \cite{3}.
Emerged from the field theoretical description for a continuum limit of the CA model \cite{6}, 
the idea of shock wave front has been pivotally inhered in the SOC
paradigm for the avalanche-type transport driven by turbulence, especially in the magnetized plasmas \cite{7}.
Also, there is an alternative which assumes a certain type of avalanche structure in random
sequence, giving the scaling laws as shown in the EG-SOC model \cite{3}.
However, they do not count exactly the role of the threshold dynamics
and the other components of SOC \cite{8}. Thus, to make the theory consistent with
the SOC concept, the avalanche transport is better to be obtained from
the continuum system with a good correspondence to all components of the SOC states.
Moreover, to enable any quantitative correlation to the physical system of
SOC behaviors, the specific model must be induced on the ground of the physical
principles in transport phenomena. 

Our aim is to introduce the equivalent continuum model ruled by the
generalized causal relation of the transport process.
By means of this approach, the mechanism of propagating instability is described
as a nonlinear wave, for the first result without losing the implication of threshold dynamics.
Showing the travelling wave consistent with the basic assumptions of SOC theory, 
we also light the unified features with the existing theories,
for instance, the EG-SOC and the shock wave description of avalanche.
Indeed, there has been many efforts to developed a well-prepared continuum model \cite{9, 10}
devoted to the role of the threshold dynamics,
or to the scaling concept, for instance, in the Landau-Ginzburg theory's point of view \cite{8}.
Remarkably, an analogy of transport events from the threshold dynamics was
implemented by the relaxation of transport coefficient \cite{9}, which made a success in
the application to some states of multi-scale complexity \cite{11,12}. 

\begin{equation}
\label{eq:1}
\begin{array} {r@{\quad~\quad}l}
 Q\left[|\nabla T| \right] = \left\{ \begin{array} {r@{\quad~\quad}l}
     \chi_{0} \rightarrow \chi_{1} & \text{~if~} |\nabla T| > g_c\\
     \chi_{1} \rightarrow \chi_{0} & \text{~if~} |\nabla T| < \beta g_c 
     \end{array} \right.\\
(\chi_1 > \chi_0 \rm{~and~} 0 < \beta < 1)
\end{array}
\end{equation}

As a heuristic model of the turbulent driven flux $\Gamma$, the relaxation of
the coefficient $\chi = -\Gamma/\nabla T$ was introduced for the amount of time to make
the critical event stabilized between the bi-stable states in Eq. \eqref{eq:1} \cite{9}.
Leading $\chi = \int_{0}^{t} \frac{Q}{\tau} e^{-\frac{t-t'}{\tau}} dt'$,
this provides an analogy of the discriminated time scale of the flipping cells
from the random deposition in CA models. 
Now, we pay attention to the 1-D transport system \eqref{eq:2},
driven by the source $S$ in the slow-driving limit \cite{3,13}.
Written in \eqref{subeq:2b}, the gradient driven flux $\Gamma$ is defined as, so to speak,
the Guyer-Krumhansl constitutive relation \cite{14,15}.
Using this relation known as a generalization of the Fick's law, 
we can show that Eq. \eqref{eq:2} is exactly equivalent to the
well-suited continuum field model introduced by Lu \cite{9}.
On the ground of the equivalence, the physical meaning
of ``weak nonlocality'' emerges from the continuum SOC system, which
imposes the finite-scale spatiotemporal memory to the response of $\Gamma\left[\nabla T\right]$
with respect to the critical events.

\begin{subequations}
\label{eq:2}
\begin{eqnarray}
 \label{subeq:2a}
\frac{\partial T}{\partial t} + \frac{\partial \Gamma}{\partial x} = S& \text{~}\\ 
 \label{subeq:2b}
\Gamma + \tau\frac{\partial \Gamma}{\partial t} 
- \tau\chi\frac{\partial^2 \Gamma}{\partial x^2}& \text{~}
= -Q \frac{\partial T}{\partial x} 
\end{eqnarray}
\end{subequations}

We can show that the solution of the constitutive relation has an
integral form, preparing the exponential tail
of temporal memory as well as the spatial correlation ruled by $g(x, t; x′, t′)$ \cite{16}.

\begin{equation}
\label{eq:3}
\Gamma = \int \frac{dt'}{\tau} \int dx' g(x,t;x',t') e^{-\frac{t-t'}{\tau}}
Q\left[|\nabla T|\right] \nabla T  
\end{equation}

where $g$ is the Green's function as of a PDE, $\frac{\partial}{\partial t}g(x,t)
-\chi(x,t)\frac{\partial^2}{\partial x^2}g(x,t) = \delta(x-x')\delta(t-t')$. 
We can interpret $g$ in the analogy of evolving probability distribution for
a particle under the dispersion process ruled by $\chi(x, t)$.
As a result, the spreading kernel $g$ brings into effect of time varying spatial
memory whose characteristic length is proportional to the square root of time interval
$i.e.$  $\left<\Delta x^2\right>^{1/2} = \sqrt{\chi\Delta t}$
when $\Delta t$ is small. As the time constant $\tau$ means the response
time of threshold driven transport, then, the spatial correlation length will
be $\xi = \sqrt{\chi\tau}$ during the single event of critical response
(transition by the critical gradient).
Such a spatiotemporal memory plays a crucial role in the overlap of avalanche events.
It is known that this leads the scale-invariance of SOC phenomena,
as it was indicated as a primary clue of the criticality in the running sandpile model \cite{6}.
Thus, to uncover the effect of spatial memory, a study discarding 
$\chi\partial^2\Gamma/\partial x^2 $ in~\eqref{eq:3} can be suggested,
but it's out of the scope of this article.
   
On the application to transport phenomena, there was a remarkable attempt
to describe the avalanche-type heat flux by derivation of a nonlinear Guyer-Krumhasl relation 
in magnetized plasmas \cite{16}.
They showed that the relation can be derived from the kinetic formulation
of the plasma transport $i.e.$ the drift-kinetic equation. However, 
such a description of heat flux didn't give a clear context to the SOC paradigm
yet \cite{16}.
Under the circumstances, we must pay attention to the generalized flux-gradient relation
associated with the continuum SOC. Based on the constitutive relation into the SOC model,
one can expect the SOC paradigm of turbulent plasma \cite{11,12}, probably including other
fluids of turbulence \cite{3},
to come down to the basic transport process with a nonlinear property to
be approximated to the critical instability. 

Such an idea of ``heat wave'' \cite{15},
in which the heat flux is described by the generalized causal relation,
is originated from the hyperbolic property given by the delayed response of $\tau$.
This implies the avalanche propagation in the wave speed $\sqrt{Q/\tau}$ ruled by
the telegraph equation \cite{15,16}. Indeed, a similar scaling of the avalanche speed in the continuum SOC
was reported as an empirical analysis without a conclusion in terms of wave \cite{12}.
Actually, we will show that the wave propagation is the effect of
nonlinear evolution related to the critical conditions, rather than just 
the characteristic speed of the telegraph equation.
On the other hand, in the numerical point of view, we can find that the hyperbolic property
of Eq. \eqref{eq:2} must be taken into account accurately to capture
the wave front of propagting avalanche \cite{17}.

\begin{equation}
\label{eq:4}
-\frac{\partial}{\partial x}Q\frac{\partial T}{\partial x} +
\tau\frac{\partial^2 T}{\partial t^2} + \frac{\partial T}{\partial t}
-\tau\frac{\partial}{\partial x}\left(\chi\frac{\partial^2 T}{\partial x \partial t}  \right)   
= S + \tau\frac{\partial S}{\partial t} 
\end{equation}

It is apparent that the SOC states in the concept of second sound (or thermal wave)
is not supported by the hyperbolicity alone,
in particular, in our case of Eq. \eqref{eq:2}.
Because, the 3$\rm^{rd}$ order term in the reduced equation
is known to smoothen the propagating front (Eq. \eqref{eq:4}) \cite{18}.
Thus, it cannot preserve the discontinuous travelling edge. 
The effect of higher order component already has been emergied in the study of
second sound related to the Jefferys-type kernel \cite{15}.

Thus, we need a rigorous approach to understand the threshold dynamics,
which is still lacking in the descriptions of avalanche propagation.
To clarify the role of the nonlinearity in $Q[|\nabla T|]$, 
we can describe the motion of instability initiated from the stepwise
perturbation at a stable condition. Assuming a linear and stationary
profile below the critical slope $g_c$, 
the dynamic evolution of perturbative deviation
can be written. Where $T(x,t) = T_0 - g_0 x + \delta T(x,t)$ ($g_0 < g_c$),

\begin{equation}
\label{eq:5}
\frac{\partial}{\partial t} \delta T = 
\frac{\partial}{\partial x}\left(\chi\frac{\partial}{\partial x} \delta T \right) -
g_0 \frac{\partial \chi}{\partial x} 
\end{equation}
  
A practical routine is applied, which is able to handle similar types of nonlinear wave,
for instance, in the ``reaction-diffusion equation'' \cite{19}.
In case of a single train of instability, which can be represented as a finite length of unstable section ($Q=\chi_1$), one 
can write $\partial \chi/\partial x$ as \eqref{eq:6}. $\Delta\chi$ is $\chi_1-\chi_0$,
$\Delta x$ is the length of the train, and $v_f$ and $v_p$ are the speed of the front and the tail respectively to the same direction.

\begin{equation}
\label{eq:6}
\frac{\partial \chi}{\partial x} = \Delta\chi\int_{0}^{t} 
\left(\delta(x-v_pt) - \delta(x-\Delta x - v_ft)\right) e^{-\frac{t-t'}{\tau}} \frac{dt'}{\tau}
\end{equation}

Such a train of unstable section gives an exponential growth and relaxation in sequence.
So, the continuum SOC can be corresponded to the EG-SOC which leads the power-laws
on the assumption of the randomness in the length of instability \cite{3}.
It is remarkable that the scale invariance in the continuum SOC will be naturally inferable
from this, if the deterministic events by the thershold are sufficiently stochastic
to make the length of instability trains pseudo-random. 
This idea can be supported by the robust scaling the continuum SOC model
reported elsewhere \cite{11}.

If $\delta T(x,t) = \tilde{T}(z)$ on the assumption of constant speed at the front 
($z = 0$) and the tail ($z = \Delta x$), an ODE for $\tilde{T}(z)$ can be written as 
$\chi\tilde{T}''=g_0\chi'-(v+\chi')\tilde{T}'$, where $z = x - vt$ ($v = v_f$ for $z = \Delta x$, 
and $v = v_p$ for $z = 0$). Assuming stationary condition, the travelling speed of the front or 
the tail can be obtained from \eqref{eq:7} as an approximation.

\begin{equation}
\label{eq:7}
\tilde{g}(z) \equiv -\tilde{T}(z) = -\frac{g_0\chi'(z)}{v + \chi'(z)}
\end{equation}

For the speed $v_f$ at the front, we can assign $z = x - v_f t$ and $z\sim\Delta x$. 
Then, $\chi'(\Delta x) = -\Delta\chi/v_f\tau$. 
At the same manner, for the speed  $v_p$ at the tail, $z = x - v_p t$ and $z\sim0$. 
Assuming $t\sim0$, $\chi'(0) = \Delta\chi/v_p \tau\left(1-v_p/v_f\exp(-\Delta x/v_f \tau)\right)$. 
Once the approximation~\eqref{eq:7} is given to describe the travelling speed with respect to the 
gradient $g = g_0+\tilde{g}$ at two points on the front or the tail, $\tilde{g}(z)$ must be consistent 
with $Q$ $i.e.$ $Q = \chi_0\rightarrow\chi_1$ at the front and $Q = \chi_1\rightarrow\chi_0$ at the tail.
Letting $g_f = g_0 + \tilde{g}(\Delta x)$ and  $g_p = g_0 + \tilde{g}(0)$, it is clear that one must 
apply the condition of $g_f > g_c > g_0$ for the front and $g_p < \beta g_c$ for the tail. 

From $g_f > g_c > g_0$, we estimated the upper and lower limit of $v_f$ as $v_c > v_f > v_0$
where $v_c = v_0\left(\frac{g_c}{g_c-g_0}\right)^{1/2}$ and $v_0 = \sqrt{\Delta\chi/\tau}$. And, it can be shown 
that $ v_f \rightarrow v_c$ when  $g_f\rightarrow g_c$ and $v_f \rightarrow v_0$ if $g_f\rightarrow \infty$.
Meanwhile, applying $g_p < \beta g_c$, a survival condition of travelling avalanche can be found, 
which can be substituted with the condition of $v_f > v_p$ for the front not to be overtaken by the tail. 
Using $\beta g_c(1-v_p/v_f e^{-\Delta x/v_f\tau})/(g_0-\beta g_c) > (v_p/v_0)^2 >0$ from~\eqref{eq:7}, 
$g_0 > \beta g_c$ came out into the open as a necessary condition for $v_f > v_p$. Attention must be paid to the obtained
 condition in \eqref{eq:8} from the analysis. 

\begin{equation}
\label{eq:8}
g_c > g_0 > \beta g_c
\end{equation}

This means that the moving instability (avalanche packet) cannot survive without {\it metastability}. 
It is exactly consistent with the assumption in the generic SOC models \cite{13}. Thus, we can state the
clear reason to introduce the metastability to the SOC theory as a survival condition
of avalanche.
Even if the parameter $\beta$ is an {\it ad hoc} component of metastability,
it has a specific physical background from the
multiscale complexity of the magnetosphere \cite{9, 11, 20}. It was interpreted as the hysteresis
by coarse grained dissipation against the change of magnetic topology \cite{21}.
On the other hand, the metastabilty has not been taken into account in the SOC paradigm 
of magnetically confined plasmas,
whereas the growth of (linear) instabilities is widely accepted as a basic concept \cite{7, 12, 22}. 
In such a context, especially in the ITG turbulence of magnetic devices \cite{23},
it will be crucial to seek for metstable states under the consideration of nonlinear effects,
which can support the anomalous transport in the fusion devices based on SOC.

The shock wave approximation is analogous with the limit of $g_0\rightarrow\beta g_c$, 
which was obtained from the Burgers' equation in the hydrodynamic limit
of coarse grained sandpile by Hwa and Kadar \cite{6}.
Because, this limit case leads $v_p\rightarrow v_f = v_0$ and
$\Delta x\rightarrow 0$ $i.e.$ $g_f\rightarrow \infty$, which is infinitesimal
with extremely sharp front. But, we have to indicate that the ``characteristic curves''
of shock wave has an opposite structure to the logic in the survival condition of the instability.
Thus, the simplification can be logically vulnerable, which was applied the perturbative field theoretic approach \cite{6},
in spite of the analogy to the limit.
On the other hand, it is remarkable that the avalanche is proportional to
the characteristic speed $v_0 = \sqrt{\Delta\chi/\tau}$. 
The upper limit of the typical avalanche speed can be estimated as $v_0 \sqrt{1/(1-\beta)}$,
because the profile may tend to be stabilized at $g=\beta g_c$ in the slow driving limit.
It is convincing to explain the previously discovered
ballastic speed in the continuum SOC model,
replacing their empirical scaling $\propto \sqrt{\chi_1/\tau}$ \cite{12}. 

We can verify the statements employing
a numerical scheme which is designed to capture the wave propagation accurately.
Based on the equation~\eqref{eq:2},  TVD-IMEX (explicit step for the hyperbolic
component and implicit for the diffusive part) scheme \cite{17} is applied
by splitting the hyperbolic component of ~\eqref{eq:2}.  
With generalized minmod limiter, 2$\rm^{nd}$ order MUSCL-type (Monotonic Upstream-Centered Scheme of Conservation Law \cite{24}) 
algorithm was implemented for the hyperbolic part using the exact Riemann soultion for a non-constant 
impedance acoustic equation in non-conservative form \cite{25}. The parameters of numerical computation were set
as $L=20$ (the system size), $N=400$ (the numebr of mesh) or $\Delta x = 0.05$, $S=0$ (without external driving), 
$g_c = 1.5$, $\chi_0 = 0.1$, $\chi_1 = 1.0$ and $\beta = 0.9$. 
As an initial condition of $T$, the stationary slope $g_0 = T_0/L$ is chosen between $g_c=1.5$ 
and $\beta g_c = 1.35$, and a small step ($\Delta T/T_0 = 1/300$) of deviation was introduced at the center of 
the profile so that the local slope of the center is $1.2 g_c$ to trigger the instability.
Paying attention to the stationary state for unperturbed propile, the boundary condition of $T$ and $\Gamma$ 
was carefully assigned, as well as the initial condition (offset) of $\Gamma$.
 
\begin{figure}
 \includegraphics[width=0.43\textwidth]{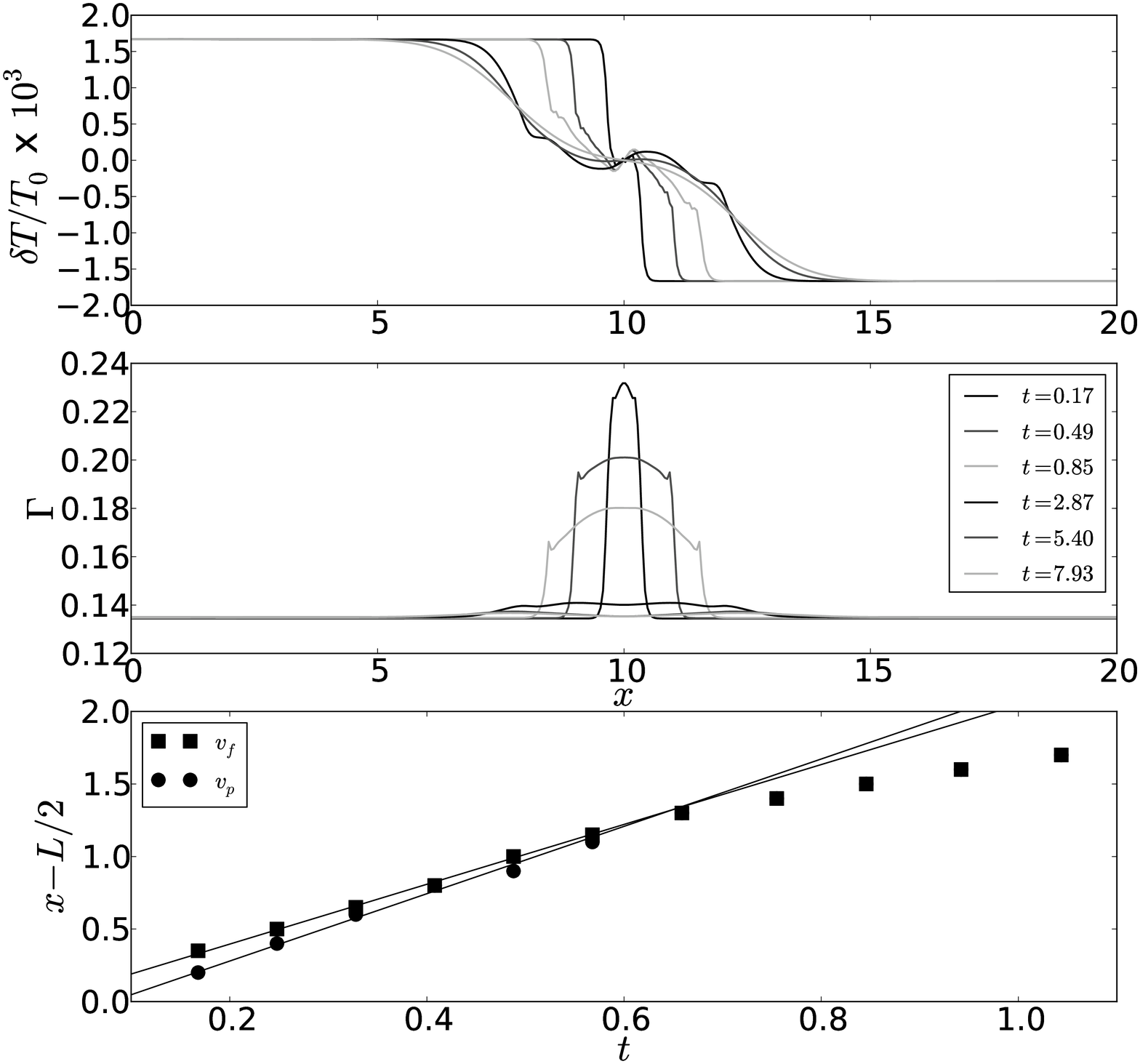}
 \caption{\label{fig:1} 
$g_0 = 1.35$ ($=0.9g_c=\beta g_c$) : the propagating condition is not satisfied.
The lower plot shows the propagation speed at the front and tail respectively ($v_f = 2.063$ and $v_p = 2.321$). 
The initial avalanche train attenuates and disappears, as $v_p > v_f$ as shown in the lower plot.}
 \end{figure}

\begin{figure}
 \includegraphics[width=0.43\textwidth]{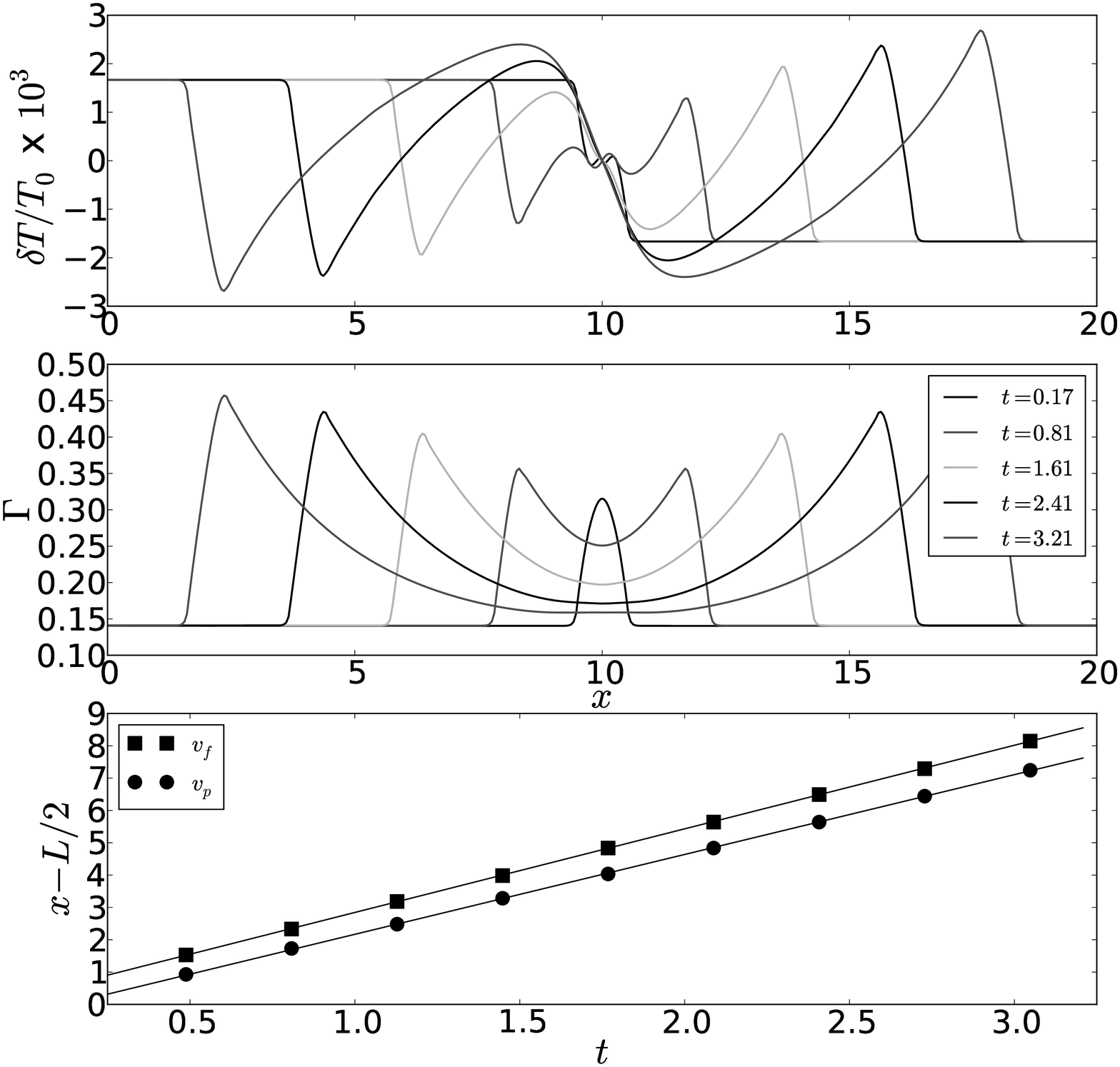}
 \caption{\label{fig:3} $g_0 = 1.41 (=0.94 g_c>\beta g_c)$ : avalanche propagation is shown, being amenable 
to the expectation $i.e.$ $v_f > v_p$ ($v_f=2.586$ and $v_p=2.469$).}
 \end{figure}

In Fig.~\ref{fig:1} $\sim$ ~\ref{fig:3}, we present the numerical simulation of the three representative cases. 
It deserves to mention that the unstble spot just produces a pair of avalanches with the ``joint-reflection symmetry''
\cite{6,7}. The numerical solution shows such a typical pattern, which is well-known as so called the ``blob-void pair'' \cite{26}, triggered by the instability at the stepwise perturbation to a stable profile.

 \begin{figure}
 \includegraphics[width=0.43\textwidth]{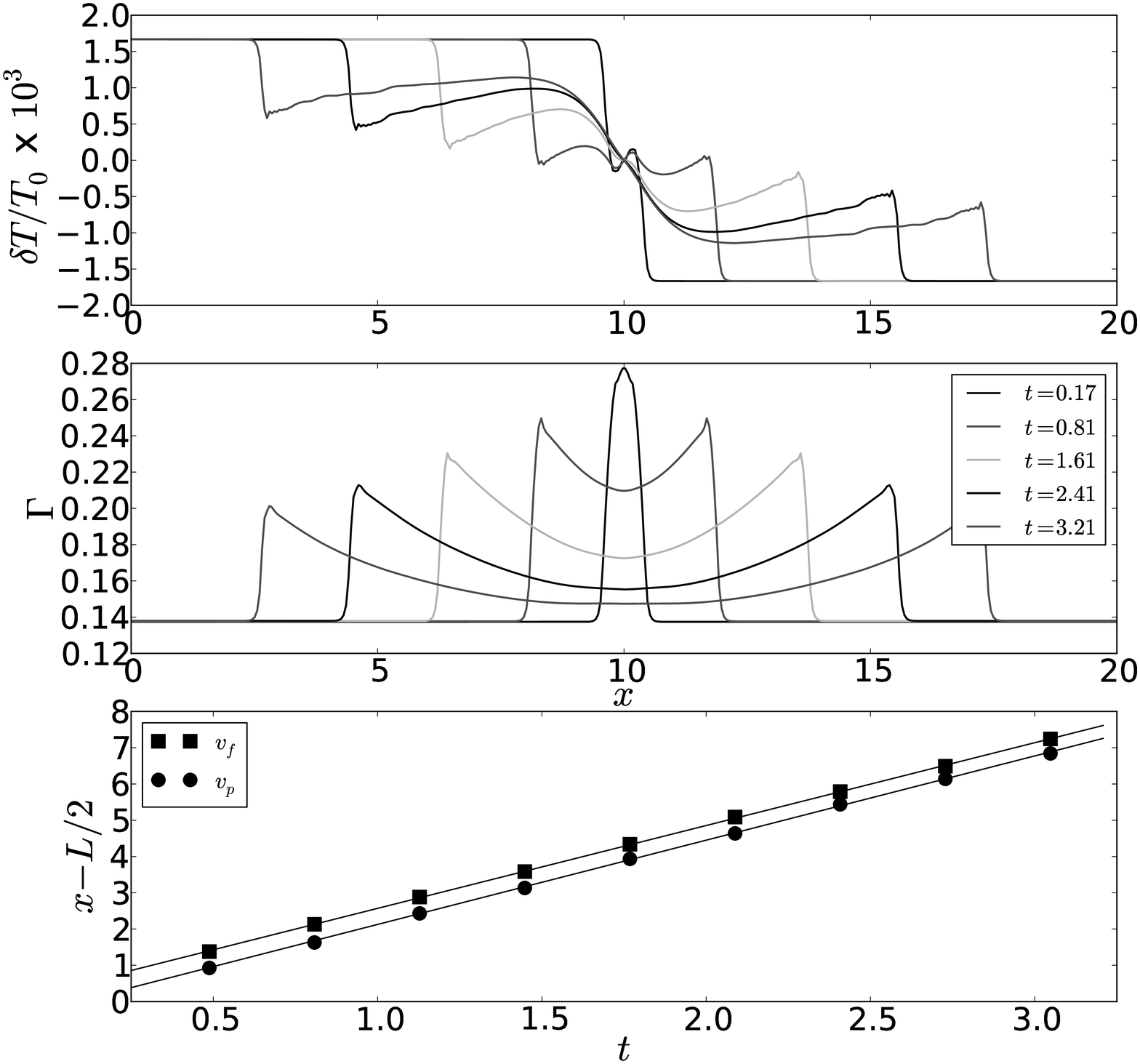}
 \caption{\label{fig:2}  
$g_0 = 1.38~(=0.92g_c>\beta g_c$) : it can be notified that this case is close to the lower limit 
of survival of avalanche train. As presented, the front and tail had almost the same speed, 
but the tail was slightly faster than front ($v_f = 2.284$ and $v_p = 2.324$). It is consistent
with decaying amplitude of $\delta T$ and $\Gamma$. This means that our analysis is somewhat
optimistic in the aspect of survival criterion.}
 \end{figure}

According to the avalanches in constant speed in \eqref{eq:7}, $v_f$ and $v_p$
can be estimated as a function of the gradients ($g_f$ at $z=\Delta x$, $g_p$ at $z=0$, and $g_0$)
and the other quantities obtained from the data, as

\begin{subequations}
\label{eq:9}
\begin{eqnarray}
 \label{subeq:9a}
v_f^{\rm calc} = v_0\sqrt{\frac{g_f}{g_f-g_0}}\\
 \label{subeq:9b}
v_p^{\rm calc} = v_0\sqrt{\frac{g_p}{g_0-g_p}\left(1-\frac{v_p^{\rm calc}}{v_f}e^{-\Delta x/v_f\tau}\right)}\\
\nonumber
\end{eqnarray}
\end{subequations}

\begin{figure}
 \includegraphics[width=0.43\textwidth]{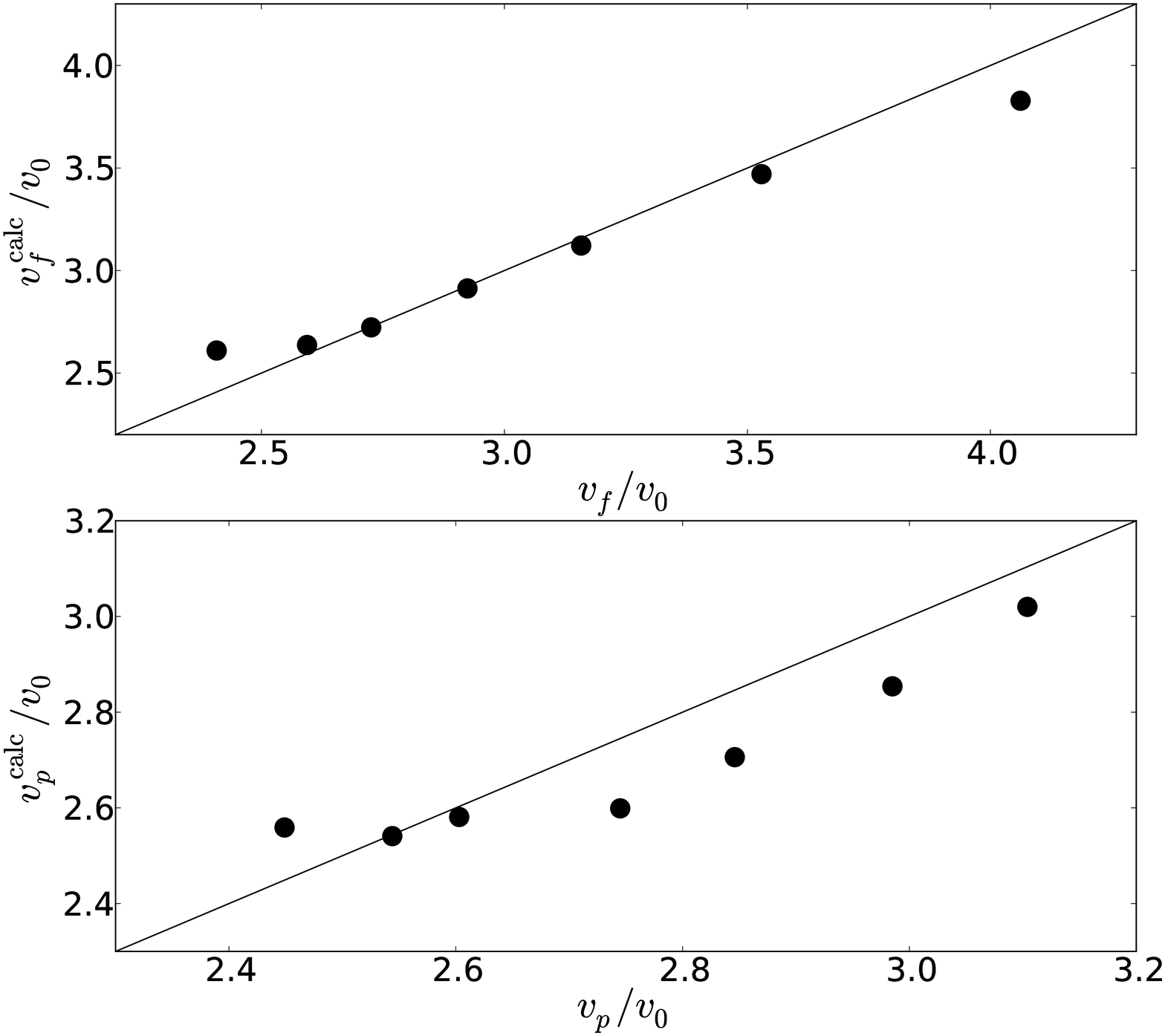} 
\caption{\label{fig:4} The speed of avalanches $v_f/v_0$ and $v_p/v_0$ comparing with the estimation $v_f^{\rm calc}/v_0$ and $v_p^{\rm calc}/v_0$ based on Eq.~\eqref{subeq:9a} and~\eqref{subeq:9b}. The stationary slopes $g_0$ are 1.38, 1.395, 1.41, 1.425, 1.44, 1.455 and 1.47 ($1.35 < g_0 < 1.5$).}
 \end{figure}

By means of Eq. \eqref{eq:9}, it is possible to test whether the simulation can validate
our description of the nonlinear wave to support the points in this letter.
The result is shown in Fig.~\ref{fig:4}, in which the numerical results is consistent
with the analysis using \eqref{eq:9} within the limit of allowed initial condition,
$i.e.$ $ 1.35 < g_0 < 1.5$.

In conclusion, as we recover the constitutive relation hidden
in the continuum SOC model, the logical basis can be secured
to describe the avalanche inferable from the transport process, 
or the kinetic description of physical system.
For the physical ground of this approach, we show
the implicated nonlinear wave in the SOC system, being
rigorously associated with the threshold dynamics.
In particular, the metastability is clearly subjected to the natural
conclusion of the wave property, which has been believed
as one of the assumptions in the SOC model.
Being inherent in the generalized causal relation, the spatiotemporal nonlocality
emerges as a critical clue of the mechanism of avalanche propagation as well as 
the scale-free nature in the continuum SOC system. 
We also refer to that the multistability in the nonlinear response should be combined with 
the nonlocality to lead the SOC-type transport as a nonlinear wave.
Thus, the result shows the continuum SOC framework
based on the physical principles of transport process, which
is supported by unified features of existing models, with a clear description of the avalanche.

Special thanks must be paid to R. Singh for interesting discussion.
This work is supported by the WCI program of National Research Foundation of Korea funded by Ministry
of Science, ICT and Future Planning of Korea [WCI 2009-001]

\end{document}